\documentclass[a4 paper, 12 pt] {article}

\begin{document}
\title{\bf{Toroidal Solitons in Nicole-type Models }}
\author{A. Wereszczy\'{n}ski  \thanks{wereszczynski@th.if.uj.edu.pl}
       \\
       \\ Institute of Physics,  Jagiellonian University,
       \\ Reymonta 4, 30-059 Krak\'{o}w, Poland}
\maketitle
\begin{abstract}
A family of modified Nicole models is introduced. We show that for
particular members of the family a topological soliton with a
non-trivial value of the Hopf index exists. The form of the
solitons as well as their energy and topological charge is
explicitly found. They appear to be identical as the so-called
eikonal knots. The relation between energy and topological charge
of the solution is also presented. Quite interesting it seems to
differ drastically from the standard Vakulenko-Kapitansky formula.
\end{abstract}
\section{\bf{Introduction}}
It is widely known that knotted solitons i.e. topological
solutions with a non-trivial value of the Hopf invariant play a
prominent role in the modern physics \cite{phys}, chemistry
\cite{chem} and biology \cite{biol}. In particular, it has been
suggested by Faddeev and Niemi \cite{niemi1} that effective
quasi-particles in the low energy limit of the quantum
gluodynamics, so-called glueballs, may be described as knotted
flux-tubes of the gauge field. In this picture a non-vanishing
value of the topological charge provides stability of
configurations and, via the Kapitansky-Vakulenko inequality
\cite{vakulenko} between energy and topological charge, fixes the
mass spectrum of glueballs. In fact, they proposed a model (the
Faddeev-Niemi model) \cite{niemi1}, \cite{cho}, where such knotted
solutions have been numerically found \cite{helmut},
\cite{battyde}, \cite{salo}. It has been also argued by many
authors that this model might be derived from the original quantum
theory \cite{shabanov}, \cite{pak}, \cite{sanchez1},
\cite{sanchez2}, \cite{vanbaal}. However, up to now, no
satisfactory proof that the Faddeev-Niemi model is the low energy
limit of the pure quantum Yang-Mill theory has been given.
\\
Unfortunately, due to the fact that the Faddeev-Niemi model
belongs to non-exactly solvable theories only numerical
\cite{helmut}, \cite{battyde}, \cite{salo} or some approximated
solutions have been obtained \cite{ward1}, \cite{ja1}. In
consequence, many problems concerning properties of the
Faddeev-Niemi hopfions have not been solved yet. Therefore, a few
models based on the same degrees of freedom and topology but
possessing analytical solutions have been constructed
\cite{nicole}, \cite{aratyn1}. For example in the
Aratyn-Ferreira-Zimerman model \cite{aratyn1} infinitely many
solitons with an arbitrary Hopf number have been found. Such toy
models gave a chance to understand connections between topological
charge and shape of a solution as well as allowed us to check the
energy-charge inequality. On the other hand, in the case of the
second widely investigated toy model i.e. the Nicole model
\cite{nicole} (it is the oldest model with explicitly found
hopfion) the spectrum of the solutions is scarcely known. Only the
simplest hopfion with $|Q_H|=1$ has been found.
\\
The main aim of the present paper is to prove that a slightly
modified Nicole models possess in their spectrum of solutions
hopfions with topological charge $Q_H=-m^2$, where $m\in Z$, and
analyze their properties like shape, energy etc.. In particular,
we are interested in checking of validity of the
Vakulenko-Kapitansky formula.
\section{\bf{Model and Solutions}}
Let us start with the following Lagrangian density
\begin{equation}
L=\frac{1}{2}\sigma (\vec{n}) (\partial_{\mu} \vec{n}
\partial^{\mu} \vec{n})^{\frac{3}{2}}, \label{model}
\end{equation}
where $\vec{n}$ is an unit, three component vector field living in
the $(3+1)$ Minkowski space-time. This model differs from the
original Nicole Lagrangian only via a function $\sigma$, which in
the case of the Nicole model is just a constant. One can see that
appearance of a non-trivial $\sigma$ function will result in the
explicit breaking of the global $O(3)$ symmetry. Models with this
property have been recently versatilely investigated
\cite{niemi2}, \cite{sanchez1}, \cite{ja2}, \cite{ferreira}. The
physical importance of such models follows from the observation
that the Faddeev-Niemi model possesses two massless Goldstone
bosons since the spontaneous $O(3)$ symmetry breaking occurs.
Thus, in order to get rid of these massless states one is forced
to implement the explicit symmetry breaking i.e. to add a new term
into Lagrangian which is not invariant under this global symmetry.
Indeed, it has been shown that in some particular patterns of the
symmetry breaking the Goldstone bosons are removed from the
spectrum of the theory and a mass gap appears \cite{wipf}.
\\
In our work the symmetry breaking function is assumed in the
following form
\begin{equation}
\sigma (\vec{n}) = \left( \frac{1+n^3}{1-n^3} \right)^{\frac{3}{2}
\left( \frac{1}{m} -1\right)} \left[
\frac{1+\frac{1+n^3}{1-n^3}}{1+\left( \frac{1+n^3}{1-n^3}
\right)^{\frac{1}{m}}} \right]^3, \label{sigma}
\end{equation}
where $m$ is an integer and positive number. Now, we take
advantage of the stereographic projection
\begin{equation}
\vec{n}= \frac{1}{1+|u|^2} ( u+u^*, -i(u-u^*), |u|^2-1).
\label{stereograf}
\end{equation}
and rewrite the Lagrangian (\ref{model}) as follow
\begin{equation}
L= \left(\frac{|u|^{\frac{1}{m}-1}}{1+|u|^{\frac{2}{m}}}\right)^3
(\partial_{\mu} u \partial^{\mu} u^*)^{\frac{3}{2}}. \label{model
st}
\end{equation}
Thus, the pertinent equation of motion reads
$$
\frac{3}{2} \partial_{\mu} \left[
\left(\frac{|u|^{\frac{1}{m}-1}}{1+|u|^{\frac{2}{m}}}\right)^3
(\partial_{\nu} u \partial^{\nu} u^*)^{\frac{1}{2}}
\partial^{\mu} u \right] - $$
\begin{equation}
(\partial_{\nu} u \partial^{\nu} u^*)^{\frac{3}{2}}
\frac{\partial}{\partial u^*} \left[
\left(\frac{|u|^{\frac{1}{m}-1}}{1+|u|^{\frac{2}{m}}}\right)^3
\right] =0. \label{all eq}
\end{equation}
Analogously as in the case of the standard Nicole Lagrangian, our
model possesses an integrable submodel defined by the additional
condition which is nothing else but the eikonal equation
\cite{adam1}, \cite{adam2}
\begin{equation}
\partial_{\mu} u \partial^{\mu} u =0. \label{eikonal1}
\end{equation}
Then, one can adopt the procedure introduced in (\cite{aratyn2})
and construct an infinite family of the conserved current.
\\
One has to remember that solutions of the integrable submodel must
obey, except upper introduced integrability condition, also
dynamical equations achieved from (\ref{all eq})
\begin{equation}
\partial_{\mu} \left[
\frac{|u|^{\frac{1}{m}-1}}{1+|u|^{\frac{2}{m}}} (\partial_{\nu} u
\partial^{\nu} u^*)^{\frac{1}{2}}
\partial^{\mu} u \right] =0. \label{eq sub1}
\end{equation}
Let us now find topological solutions of the integrable submodel
(\ref{eikonal1})-(\ref{eq sub1}). The first step is to introduce
the toroidal coordinates
$$ x=\frac{\tilde{a}}{q} \sinh \eta \cos \phi , $$
$$ y=\frac{\tilde{a}}{q} \sinh \eta \sin \phi , $$
\begin{equation}
z=\frac{\tilde{a}}{q} \sin \xi ,\label{tor_coord}
\end{equation}
where $q=\cosh \eta -\cos \xi$ and $\tilde{a}$ is a dimensional
constant which fixes the scale in the coordinates. Moreover, the
solution is assumed to have the following form (see
\cite{aratyn1})
\begin{equation}
u(\eta, \xi, \phi )= f(\eta) e^{im(\xi+\phi)}, \label{anzatz}
\end{equation}
where unknown function $f$ is yet to be determined. It is a well
known fact \cite{aratyn1}, \cite{adam1}, \cite{ja2} that for
smooth functions $f$ such that $f(0)=\infty$ and $f(\infty)=0$ map
(\ref{anzatz}) corresponds to a non-vanishing value of the
topological charge. Indeed, one can get that
\begin{equation}
Q_H=-m^2. \label{hopf}
\end{equation}
Inserting Ansatz (\ref{anzatz}) into equation (\ref{eq sub1}) we
obtain
$$
\partial_{\eta} \left[ \frac{f^{\frac{1-m}{m}}}{1+f^{\frac{2}{m}}}
\left( f'^2_{\eta} +m^2 \frac{\cosh^2 \eta}{\sinh^2 \eta} f^2
\right)^{\frac{1}{2}} f'_{\eta} \right]- $$ $$ m^2 \frac{\cosh^2
\eta}{\sinh^2 \eta} \frac{f^{\frac{1-m}{m}}}{1+f^{\frac{2}{m}}} f
\left( f'^2_{\eta} +m^2 \frac{\cosh^2 \eta}{\sinh^2 \eta} f^2
\right)^{\frac{1}{2}} + $$
\begin{equation}
+ \frac{f^{\frac{1-m}{m}}}{1+f^{\frac{2}{m}}} \left( f'^2_{\eta}
+m^2 \frac{\cosh^2 \eta}{\sinh^2 \eta} f^2\right)^{\frac{1}{2}}
f'_{\eta} \frac{\cosh \eta}{\sinh \eta} =0. \label{eq mot1}
\end{equation}
After some calculations, it can be reduced to the more compact
form
\begin{equation}
\partial_{\mu} \ln \left[\frac{f^{\frac{1-m}{m}}}{1+f^{\frac{2}{m}}}
\left( f'^2_{\eta} +m^2 \frac{\cosh^2 \eta}{\sinh^2 \eta} f^2
\right)^{\frac{1}{2}} |f'_{\eta}| \right] +m^2\frac{\cosh^2
\eta}{\sinh^2 \eta} \frac{f}{|f'|} + \frac{\cosh \eta}{\sinh
\eta}=0. \label{eq mot2}
\end{equation}
On the other hand, our submodel is defined not only by the
dynamical field equation (\ref{eq sub1}) but also by the constrain
(\ref{eikonal1}), which in the case of upper introduced Ansatz
takes the following form
\begin{equation}
f'^2_{\eta}=m^2 \frac{\cosh^2 \eta}{\sinh^2 \eta} f^2.
\label{eikonal2}
\end{equation}
Thus, the dynamical equation can be simplified
\begin{equation}
\partial_{\eta} \ln \left[ \frac{f^{\frac{1-m}{m}}}{1+f^{\frac{2}{m}}}
f'^2\right]=-(m+1) \partial_{\eta} \ln \sinh \eta. \label{eq mot3}
\end{equation}
Applying once again constrain (\ref{eikonal2}) we find that
\begin{equation}
\partial_{\eta} \ln \left[ \frac{\cosh^2 \eta}{\sinh^2 \eta}  \frac{f^{\frac{1+m}{m}}}{1+f^{\frac{2}{m}}}
\right]=-(m+1) \partial_{\eta} \ln \sinh \eta. \label{eq mot4}
\end{equation}
This differential equation can be easily solved and in consequence
we derive an algebraic equation for $f$
\begin{equation}
\frac{\cosh^2 \eta}{\sinh^2 \eta}
\frac{f^{\frac{1+m}{m}}}{1+f^{\frac{2}{m}}}= \left( \frac{1}{\sinh
\eta} \right)^{m+1}. \label{eq mot5}
\end{equation}
The solution of this equation reads
\begin{equation}
f(\eta ) =\frac{1}{\sinh^m \eta}. \label{sol shape}
\end{equation}
One can check that it solves our constrain (\ref{eikonal2}) as
well. Thus, Ansatz (\ref{anzatz}) where the shape function i.e.
function $f$ takes the upper obtained form (\ref{sol shape}) is a
static, topologically non-trivial solution of the integrable
submodels. One can immediately notice that for $m=1$ the
well-known unit charge hopfion which is a solution to the standard
Nicole model is reproduced.
\\
It should be underlined that every exact solution (\ref{sol
shape}), label by $m \in N$, refers to the different modified
Nicole Lagrangian. We have proved that any model (\ref{model})
with $m \in N$ possesses a topological solution with $Q_H=-m^2$.
It is unlikely the Aratyn-Ferreira-Zimerman model where an
infinite family of hopfions has been found. Nonetheless, our
calculation shows that also in the framework of the modified
Nicole models some exact hopfions with higher than one topological
charge can be constructed.
\\
Let us now compute the corresponding energy. Using the
stereographic projection we derive
\begin{equation}
E=\int d^3 x
\left(\frac{|u|^{\frac{1}{m}-1}}{1+|u|^{\frac{2}{m}}}\right)^3
(\partial_{i} u \partial_{i} u^*)^{\frac{3}{2}}.
\end{equation}
Taking into account the form of the Ansatz one can rewrite this
expression as follows
\begin{equation}
E=(2\pi)^2 \int_0^{\infty} d\eta \sinh \eta \left(
\frac{f^{\frac{1-m}{m}}}{1+f^{\frac{2}{m}}}\right)^3 \left(
f'^2_{\eta} +m^2 \frac{\cosh^2 \eta}{\sinh^2 \eta} f^2
\right)^{\frac{3}{2}}.
\end{equation}
Finally, inserting our solution we obtain that
\begin{equation}
E=(2\pi)^2 2^{\frac{3}{2}} m^3 \int_0^{\infty} \frac{\sinh
\eta}{\cosh^3 \eta} = \sqrt{2} (2\pi)^2 m^3.
\end{equation}
Quite interesting, the energy of the hopfion is related with its
topological charge by the following relation
\begin{equation}
E=\sqrt{2} (2\pi)^2 |Q_H|^{\frac{3}{2}}, \label{eq charge}
\end{equation}
which differs considerable from the standard Vakulenko-Kapitansky
formula. Vakulenko and Kapitansky proved that in the case of the
Faddeev-Niemi model energy of the solution is bounded from below
by the corresponding topological charge. Namely,
\begin{equation}
E \geq C |Q_H|^{\frac{3}{4}}, \label{kapitanski}
\end{equation}
where $C$ is a numerical constant. Recently, new results
concerning upper bound have been presented in \cite{lin}. It has
been also shown that asymptotically for large topological charge
energy is proportional to $|Q|^{3/4}$. Moreover, it was checked by
direct calculations that this relation is valid for all known
solutions of the Aratyn-Ferreira-Zimerman model \cite{aratyn1} as
well as its generalizations \cite{ja2}, \cite{ja3}. Indeed, the
energy grows proportional to $|Q|^{3/4}$.
\\
Here, for the modified Nicole models, such sublinear behavior is
not longer held. Indeed, the exponent characterizing the
dependence on the topological index is bigger than one and reads
$\frac{3}{2}$. Of course, the Vakulenko-Kapi\-tan\-sky inequality
is valid since $E \geq C |Q|^{3/2} \geq C |Q|^{3/4}$. Nonetheless,
the different value of the exponent can result in the modification
of the interaction between hopfions. Instead of the standard
clustering phenomena (a separated multi-soliton configuration
tends to form a clustered, really knotted state) one should rather
expect splitting i.e. decay a soliton with high topological
charges into unknots with unit Hopf index.
\\
One can notice that there may be a trivial solution to this
unexpected relation between energy and the topological index.
Namely, presented solitons are not the energy minimums in the
fixed topological sector. Then there may exist less energy
solutions which would saturate the Vakulenko-Kapitansky formula.
Such possibility is also interesting as it suggests that stable
configurations could be given not by obtained here unknots but by
really knotted solitons. Due to the fact that such a property is
observed in the Faddeev-Niemi model \cite{battyde}, \cite{salo},
it would indicate that the Nicole toy model is much more relevant
to investigation of hopfions than the Aratyn-Ferreira-Zimerman
model.
\\
However, it must be stressed once again that we do not know
whether all hopfions corresponding to any of the modified models
follow relation (\ref{eq charge}) since only one hopfion for each
modified model has been obtained. Thus, as far as no solutions
with other values of the topological charge will be found, our
energy-charge relation has to be treated only as a conjecture.
\section{\bf{Conclusions}}
In the present work, a modification of the Nicole Lag\-ran\-gian
has been considered. For each member of the family of the modified
models (\ref{model}) (label by an integer and positive parameter
$m$) a topological solution with $Q_H=-m^2$ has been found. Let us
shortly summarize the obtained results.
\\
First of all, we have shown that all solitons are unknots that is
surfaces corresponding to constant values of the unit, vector
field $\vec{n}$ are toruses. It resembles situation known from
Aratyn-Ferreira-Zimerman model. This fact can be treated as a
disadvantage of the toy models since Faddeev-Niemi hopfions are
really knotted objects without toroidal symmetry.
\\
However, the most important observation we have made concerns the
energy-charge inequality. As we have discussed it before, there
are some arguments indicating that energy of the hopfions for each
of the modified Nicole model is proportional to $Q_H^{3/2}$ rather
than $Q_H^{3/4}$ as one could expected from the
Vakulenko-Kapitansky inequality.
\\
Undoubtedly, further studies are needed. For example, the validity
of this conjecture should be checked. We would like to address
this problem in the forthcoming paper.
\\ \\
This work is partially supported by Foundation for Polish Science
FNP and ESF "COSLAB" programme.

\end{document}